\documentclass[%
 reprint,
 superscriptaddress,
 amsmath,amssymb,
 aps,
prb,
]{revtex4-2}

\usepackage{graphicx}
\usepackage{dcolumn}
\usepackage{bm}
\usepackage{hyperref}

\begin{document}


\title{Spinon excitations and spin correlations in the one-dimensional quantum magnet $\beta$-VOSO$_4$ probed by Raman spectroscopy}

\author{Dirk Wulferding}
 \email{d.wulferding@sejong.ac.kr}
 \affiliation{Department of Physics and Astronomy, Sejong University, Seoul 05006, Republic of Korea}

\author{Diana Lucia Quintero-Castro}
 \affiliation{Department of Mathematics and Physics, University of Stavanger, 4036 Stavanger, Norway}
 \affiliation{PSI Center for Neutron and Muon Sciences, 5232 Villigen PSI, Switzerland}

\author{Pontus Laurell}
 \affiliation{Department of Physics and Astronomy, University of Tennessee, Knoxville, TN 37996, USA}
 \affiliation{Department of Physics and Astronomy, University of Missouri, Columbia, MO 65211, USA}
 \affiliation{Materials Science and Engineering Institute, University of Missouri, Columbia, MO 65211, USA}

\author{Gonzalo Alvarez}
 \affiliation{Computational Sciences \& Engineering Division and Center for Nanophase Materials Sciences, Oak Ridge National Laboratory, Oak Ridge, TN 37831, USA}

\author{Elbio Dagotto}
 \affiliation{Department of Physics and Astronomy, University of Tennessee, Knoxville, TN 37996, USA}
 \affiliation{Materials Science and Technology Division, Oak Ridge National Laboratory, Oak Ridge, TN 37831, USA}

\author{Kwang-Yong Choi}
 \email{choisky99@skku.edu} 
 \affiliation{Department of Physics, Sungkyunkwan University, Suwon 16419, Republic of Korea}

\date{\today}

\begin{abstract}

Fractionalized excitations such as spinons and anyons have emerged as a central theme in condensed matter physics with broad implications for superconductivity, quantum statistics, and quantum computation. The nearly ideal one-dimensional $S=1/2$ system $\beta$-VOSO$_4$ without long-range order down to 85 mK provides a promising platform to experimentally explore such fractionalized excitations. Here, we employ Raman spectroscopy to probe magnetic excitations and the evolution of spin correlations in $\beta$-VOSO$_4$. Spinon signatures are found along the chain direction, evidenced by a broad, gapless scattering continuum at low temperatures. The temperature dependence of the spinon spectral weight aligns considerably with numerical density matrix renormalization group calculations. By comparing the experimental spinon spectral weight with calculated results and evaluating the associated quantum Fisher information (QFI) therefrom, we observe a steep increase in QFI upon cooling, indicating rapidly growing correlation lengths. Our study showcases QFI as a probe of spin correlations in quantum magnets.

\end{abstract}

\maketitle

\section{\label{intro}Introduction}

Quantum entanglement is one of the major concepts in modern physics, with implications ranging from string theory to encrypted communication approaches~\cite{horodecki-09} to the emergence of spacetime~\cite{swingle-18}. Despite its fundamental importance, it remains monumentally challenging to experimentally probe and verify the existence of entangled quantum states in condensed matter systems. One promising platform for exploring quantum entanglement is quantum spin liquids, a quantum liquid-like state of magnetic materials, where competing interactions frustrate conventional long-range order. Instead of forming static magnetic order, these highly dynamic ground states are governed by quantum fluctuations and entangled spins~\cite{balents-10}. However, probing the entanglement between spins directly proves to be challenging as well. Alternatively, quantum entanglement witnesses are sought out that can verify the presence of quantum entanglement in a quantum system without necessarily measuring the entangled state itself. Within such a scheme, whether a quantum system is entangled can be assessed by partial information about its state through specific measurements. However, even if a specific witness fails to detect entanglement, it does not necessarily mean the absence of quantum entanglement. As such, multiple entanglement witnesses based on distinct observable quantities should be devised to ensure entanglement.

Recently, a promising approach was introduced to experimentally quantify the degree of entanglement in quantum spin systems by combining density matrix renormalization group (DMRG) technique and inelastic neutron scattering (INS) methods~\cite{hauke-16, scheie-21, laurell-24, scheie-25}. The latter technique is especially powerful as it allows us to access the magnetic spectral weight associated with the quantum spin state in absolute units, which can be directly related to the quantum Fisher information (QFI) and, thereby, to the depth of entanglement. On the other hand, novel quantum magnet materials are oftentimes available only in sub-mm sample sizes, insufficient for INS experiments, and the generally limited beamtime at large facilities significantly limits the dataset that may be acquired. Furthermore, it may not always be straightforward in INS to effectively separate the spin-only part from other contributions (e.g., phonons) that may overlap energetically. These drawbacks call for alternative, complementary spectroscopic tools to measure spin correlation functions and assess entanglement as a function of various external tuning parameters. Magnetic Raman scattering can directly access fractionalized spinons and thereby help to overcome these limitations: while first-order scattering processes are probing excitations around the Brillouin zone center, higher-order momentum- and spin-conserving scattering processes in zigzag antiferromagnets can measure contributions from the entire Brillouin zone (most prominent contributions from the zone boundary are enabled in the presence of zone-folding)~\cite{sokolik-22, hakani-24, pawbake-25}. Thus we may gain access to the observable $O_\mathrm{R} \sim \langle S_i S_j S_k S_l \rangle$, conveying information about four-spin correlation functions between spins located at sites $i, j, k,$ and $l$~\cite{fedders-77, devereaux-07}. By inferring entanglement witness from a magnetic Raman response function, we can obtain deeper insights into the spin correlations of quantum spin systems. Among various spin systems, $S=1/2$ spin chains serve as excellent testbeds due to their analytical solvability and the availability of advanced numerical techniques.

Here, we employ Raman spectroscopy -- a simple, cost-effective table-top technique suitable for $\mu$m-sized crystals or monolayer materials -- to directly probe fractionalized spinon excitations in the quantum spin chain $\beta$-VOSO$_4$~\cite{quintero-castro-22}. This quasi-1D $S=1/2$ compound evidences no long-range magnetic order down to at least $T = 85$ mK. Both the Raman spectral response from four-spin correlations and its temperature dependence give important clues about the spin correlations of a Luttinger spin liquid. By comparing the spinon spectral weight to DMRG results, as well as by extracting QFI therefrom, we show that the thermal evolution closely scales onto the DMRG results. This scaling allows us to quantify the growth of spin correlations from $n_\mathrm{e} = 1$ to $n_\mathrm{e} \ge 3$ with decreasing temperature from a paramagnetic to a liquid-like state.

\section{\label{results}Results}

\subsection{Raman-spectroscopic fingerprints of spinon excitations}

$\beta$-VOSO$_4$ crystallizes in the orthorhombic structure $Pnma$ (space group 62). It forms chains of tilted VO$_6$ octahedra running along the $a$-axis, which are interconnected within the $bc$-plane by SO$_4$ tetrahedra [see Fig. 1(a)]. The atoms are located at the following Wyckoff positions: V($4c$), S($4c$), O1($4c$), O2($4c$), O3($4c$), and O4($8d$). In this irreducible configuration, 42 Raman-active phonons are allowed: 13 $A_\mathrm{g}$ + 8 $B_\mathrm{1g}$ + 13 $B_\mathrm{2g}$ + 8 $B_\mathrm{3g}$~\cite{aroyo-1, aroyo-2, aroyo-3}. Their corresponding Raman tensors are:

\begin{center}

\begin{widetext} \mbox{$A_\mathrm{g}$=$\begin{pmatrix} a & 0 & 0\\ 0 & b & 0\\ 0 & 0 & c\\
\end{pmatrix}$

, $B_\mathrm{1g}$=$\begin{pmatrix} 0 & d & 0\\ d & 0 & 0\\ 0 & 0 & 0\\ \end{pmatrix}$

, $B_\mathrm{2g}$=$\begin{pmatrix} 0 & 0 & e\\ 0 & 0 & 0\\ e & 0 & 0\\ \end{pmatrix}$

, $B_\mathrm{3g}$=$\begin{pmatrix} 0 & 0 & 0\\ 0 & 0 & f\\ 0 & f & 0\\ \end{pmatrix}$}

\end{widetext}
\end{center}

\begin{figure*}
\includegraphics[width=16cm]{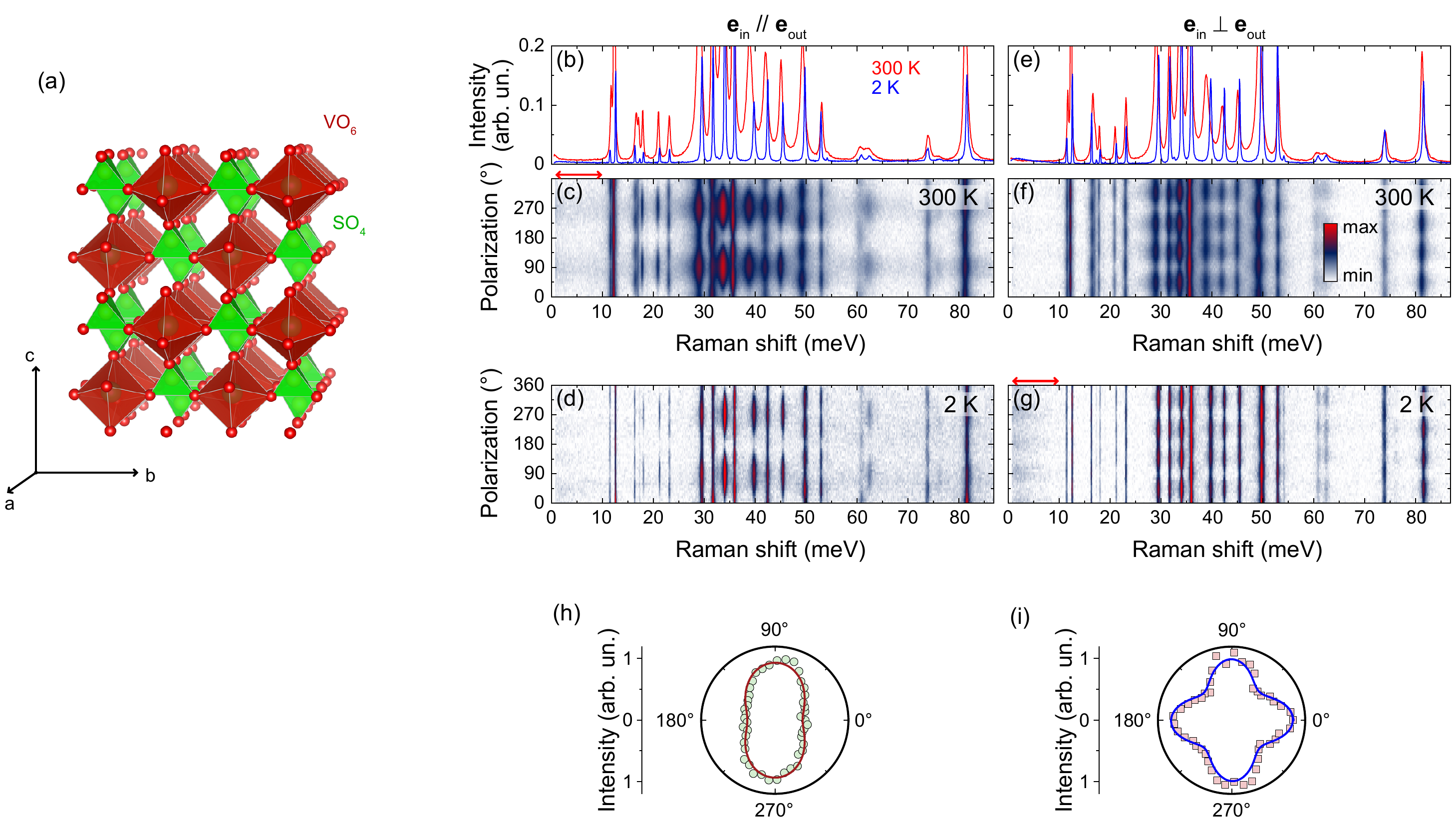}
\caption{\label{fig:polar} \textbf{Structure and symmetry properties of $\beta$-VOSO$_4$.} \textbf{a)} Quasi-1D chains of VO$_6$ octahedra (red) interconnected by SO$_4$ tetrahedra (green). \textbf{b)} Comparison of phonon spectra at $T = 300$ K and 2 K measured in parallel polarization. \textbf{c)} and \textbf{d)}: Color-contour plots of the as-measured Raman scattering intensity recorded with $\textbf{e}_{\mathrm{in}} \parallel \textbf{e}_{\mathrm{out}}$ as a function of in-plane light polarization taken at $T = 300$ K and at $T = 2$ K, respectively. \textbf{e)} Comparison of phonon spectra at $T = 300$ K and 2 K measured in crossed polarization. \textbf{f)} and \textbf{g)}: Color-contour plots of the as-measured Raman scattering intensity recorded with $\textbf{e}_{\mathrm{in}} \perp \textbf{e}_{\mathrm{out}}$ as a function of in-plane light polarization taken at $T = 300$ K and at $T = 2$ K, respectively. \textbf{h)} Polar plot of quasi-elastic scattering intensity measured at $T = 300$ K in parallel polarization. \textbf{i)} Polar plot of integrated intensity of the continuum measured at $T = 2$ K in crossed polarization. The solid lines represent fits to the data. The red arrows in panels \textbf{c)} and \textbf{g)} mark the energy range over which the scattering intensities shown in the polar plot panels were integrated.}
\end{figure*}

In Figs. 1(b)-1(g), we show polarization-resolved Raman scattering results. These measurements are carried out at $T = 300$ K [Figs. 1(c),(f)] and at $T = 2$ K [Figs. 1(d),(g)]. Numerous sharp phonon modes are clearly observed within the displayed energy range, with additional Raman-active phonons residing at higher energies. Besides the well-resolved phonons, we observe two additional scattering contributions: (i) A weak quasi-elastic scattering (QES), i.e., a scattering contribution with finite linewidth centered at $E=0$. It is observed dominantly in parallel polarization and at high temperatures, following a two-fold symmetry marked by a red arrow above Fig. 1(c). (ii) A shallow, broad scattering continuum without any clear structure, primarily observed in crossed polarization and at low temperatures with a four-fold symmetry. This continuum emerges below about 10 meV and exists down to the lowest energy, i.e., it appears to be gapless [see the red arrow in Fig. 1(g)]. In the polar plots of Figs. 1(h) and 1(i), we show the integrated intensities of the QES (green circles) and the continuum (red squares) as a function of polarization. The solid lines are fits to the data based on the Raman tensors. The high-temperature QES behavior exhibits characteristics consistent with $A_\mathrm{g}$ symmetry, while the low-temperature continuum adheres to $B_\mathrm{g}$ symmetry. This is rationalized by the different origins of these two signals: QES arises from (thermally-driven) fluctuations of spin energy density within a spin chain, and thus shows a two-fold symmetry. Conversely, the spinon continuum involves spinon-antispinon excitations, with interchain interactions modifying the selection rule accordingly.

\begin{figure}
\includegraphics[width=8cm]{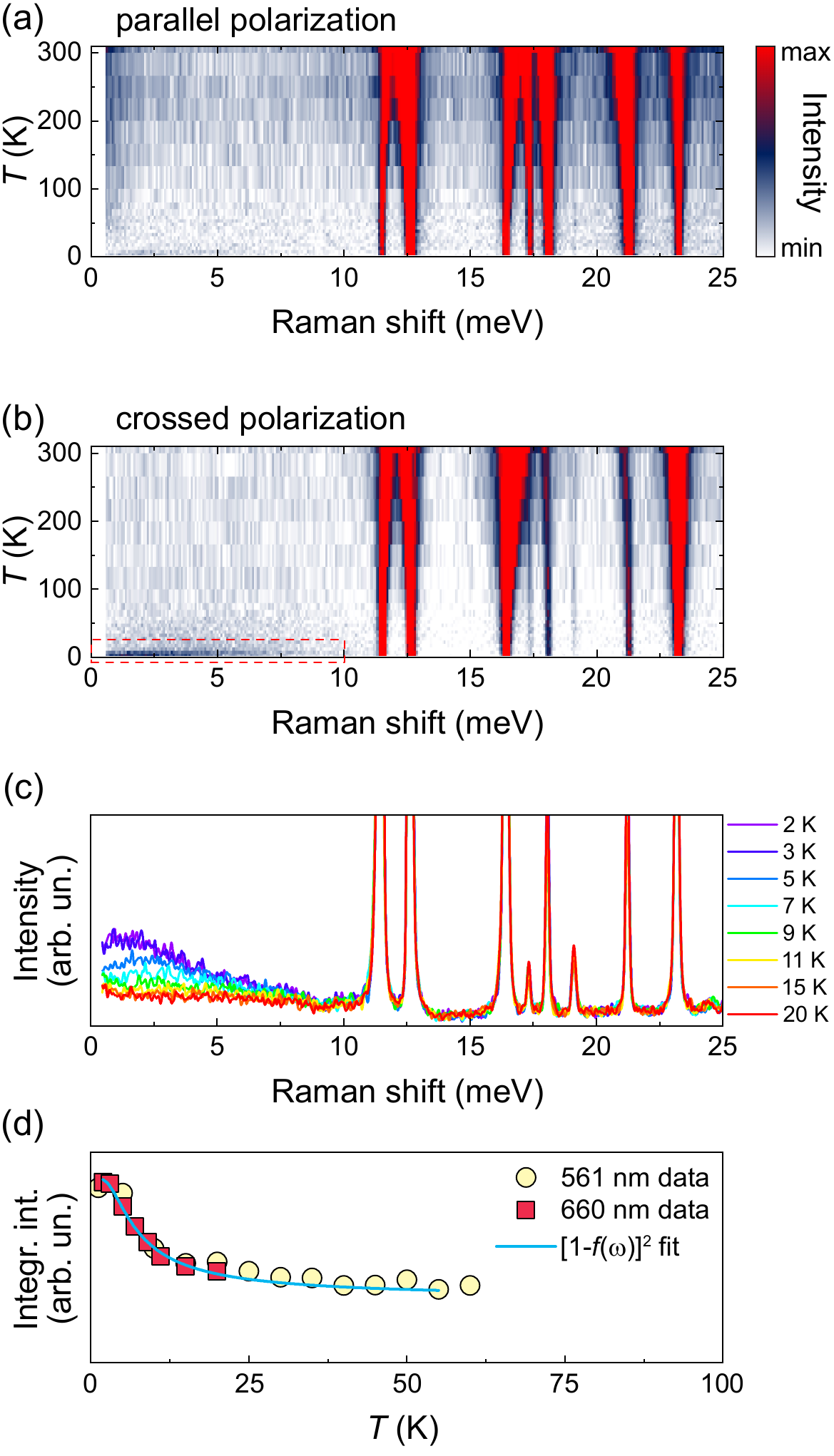}
\caption{\label{fig:Tdep} \textbf{Thermal evolution of spin correlations in $\beta$-VOSO$_4$.} \textbf{a)} and \textbf{b)} Color-contour plots of the as-measured Raman scattering intensity as a function of temperature, focusing on the low-energy regime up to 25 meV. Spectra measured in parallel polarization and in crossed polarization, respectively. The dashed red box guides to the temperature-energy range in which spinon excitations are observed. \textbf{c)} As-measured Raman spectra recorded in crossed polarization with a focus on low energies and low temperatures. The spinon continuum, extending up to $\approx$ 10 meV, is strongly diminished with increasing temperatures. \textbf{d)} Scattering intensity of the spinon continuum, integrated from 0.6 meV to 10 meV, as a function of temperature. The symbols correspond to data obtained with a $\lambda = 561$ nm laser (yellow circles) and a $\lambda = 660$ nm laser (red squares). The solid blue line is a fit to the data based on a fermionic model ([$1-f(\omega)^2$]; see the main text for details).}
\end{figure}

To glimpse into the origin of the continuum and the QES, we now focus on the temperature dependence of excitations residing at low energies (0.6 meV -- 25 meV) measured in both parallel [Fig. 2(a)] and crossed polarization [Fig. 2(b)]. Overall, the phonons in this energy window do not show evidence for any dramatic dynamics that could be related to a structural phase transition (e.g., a spin-Peierls transition) across the investigated temperature range. There is, however, ample evidence for spin-phonon coupling, as most of the phonons show subtle anomalies in linewidth and energy as the system transitions from the paramagnetic spin-gas state into the correlated spin-entangled phase (D. L. Quintero-Castro, personal communication). Contrary to the phonons, the broad low-energy continuum has a dramatic temperature dependence as it rapidly melts away with increasing temperature [see the dashed red box in Fig. 2(b)]. While the continuum is largely absent in parallel polarization, the subtle QES emerges once temperatures exceed $\approx 60$ K [Fig. 2(a)]. QES is a common observation in low-dimensional spin systems and marks the cross-over from the highly entangled into the uncorrelated spin-gas phase~\cite{wulferding-19}. Indeed, phonon frequencies experience a sudden renormalization together with a discontinuous change in linewidth around 60 K, a fingerprint of spin-phonon coupling, and therefore evidence for enhanced spin-spin correlations below 60 K (D. L. Quintero-Castro, personal communication).

In Fig. 2(c), we plot individual as-measured Raman spectra in crossed polarization with a focus on low energies at a few selected temperatures. The broad continuum is clearly observed in crossed polarization and at the lowest measured temperature $T = 2$ K. As the temperature increases up to 20 K, its intensity is rapidly reduced and replaced by a temperature-independent plateau. In contrast, in parallel polarization the broad continuum is largely suppressed. Next we carefully trace the temperature dependence of the continuum's integrated intensity [$I(T)$] up to 10 meV [see Fig. 2(d)]. For our analysis, we use datasets obtained with $\lambda = 660$ nm (red squares) and $\lambda = 561$ nm (yellow circles) lasers. We find a consistent behavior between these datasets. Therefore, we can rule out the possibility of unwanted resonance processes obscuring the low-energy spectral range. The integrated intensity of the continuum peaks at the lowest temperatures, followed by a rapid intensity drop and a nearly temperature-independent plateau above 20 K. We can fit this behavior well by a fermionic response, $I(T) \sim (1-f)^2$, with $f$ being the Fermi-Dirac function $f=1/[1+\mathrm{e}^{\beta \omega_\mathrm{F}}]$~\cite{nasu-16, sandilands-15, glamazda-16}, $\beta$ the reduced temperature, and $\omega_\mathrm{F} \approx 11$ K the characteristic energy of involved fermionic degrees of freedom. This approach has been highly successful in describing fractionalization of quasiparticles into fermionic degrees of freedom in related quantum magnets, such as Kitaev honeycomb magnets with Majorana-fermionic excitations~\cite{sandilands-15, glamazda-16}, and other frustrated, low-dimensional quantum magnets with fractionalized spinon excitations. The shape of the low-energy spectral weight is also characteristic for such kind of systems: the continuum shows no evidence of any gap (down to at least 0.6 meV), and it forms a broad maximum followed by a long, gradual drop-off to higher energies up to, or beyond, $\approx 10$ meV (i.e., $E_{\mathrm{cutoff}} > J$). All these characteristics are commonly found in quantum spin liquid candidates~\cite{wulferding-10, wulferding-20, gnezdilov-21, sokolik-22}, while the drop-off energy is comparable to INS experiments~\cite{quintero-castro-22}, where continua of 2- and 4-spinon excitations were detected up to at least 7 meV. We can therefore confidently assign the broad low-energy scattering signal observed at low temperatures to a continuum of spinon excitations, i.e., a hallmark of spin fractionalization in Heisenberg chains. In the following, we will model the dynamical spin structure factor via DMRG at finite temperatures and compare the results to the experimentally extracted spinon continuum. This analysis will allow us to evaluate QFI and quantify quantum entanglement.

\subsection{DMRG modeling}

Following its initial description and parametrization~\cite{quintero-castro-22}, we model $\beta$-VOSO$_4$ as an isotropic $S = 1/2$ Heisenberg antiferromagnetic chain with the spin Hamiltonian

\begin{equation}
H = J \sum^{L-1}_{j=0}{\textbf{S}_j \cdot \textbf{S}_{j+1}},	\label{eq:hafc}
\end{equation}
where $J \approx 3.83(2)$ meV is the nearest-neighbor intrachain coupling strength, $j$ is a site index along the chain, $L$ is the length of the chain, and $\textbf{S}_j = \vec{\sigma}_j/2$ are spin-1/2 operators, with $\vec{\sigma}_j$ the vector of Pauli matrices. We emphasize that the exchange coupling $J$ is comparable to $J_\mathrm{R} = 3.2$ meV, estimated from the high-energy cutoff of the spinon continuum seen in Raman scattering via $E_{\mathrm{cutoff}} = \pi J_\mathrm{R} = 10$ meV~\cite{loosdrecht-96}. A random phase approximation fit was performed previously~\cite{quintero-castro-22} to estimate the interchain coupling strength $J_\perp$ in a $J - J_\perp$ model, yielding $\vert J_\perp \vert /J \approx 0.05$. This ratio is comparable to the corresponding values in the Heisenberg quantum spin chain compounds KCuF$_3$, which is described well with $\vert J_{\perp} \vert /J \approx 0.047$~\cite{lake-05, scheie-24}, and Sr$_2$V$_3$O$_9$ with an estimated $\vert J_{\perp} \vert /J \approx 0.023$ \cite{gao-23}. 

We note two potential sources of uncertainty that could affect the estimated ratio for $\beta$-VOSO$_4$: (i) the fit was performed on powder data, and (ii) there could potentially be a small frustrating next-nearest neighbor coupling $J_2>0$ that was not taken into account in the fit \cite{quintero-castro-22}. Nevertheless, the fit does suggest the interchain and $J_2$ couplings are energetically small, such that the dynamical spin structure is expected to be well approximated by Eq. \eqref{eq:hafc}. In particular, since small couplings mainly modify the low-energy scattering, which is suppressed by the \texttt{tanh} filter function in the QFI integral [Eq. (2) below], $\beta$-VOSO$_4$ would be expected to have entanglement depths and scaling of QFI with temperature that is comparable to KCuF$_3$ \cite{scheie-21, laurell-24, scheie-25}.

The finite-temperature dynamical spin structure factor of Eq. \eqref{eq:hafc} was studied using the DMRG technique~\cite{white-92, white-93} using the DMRG++ software~\cite{alvarez-09}. These calculations closely follow those outlined in earlier works~\cite{scheie-21, gao-23} and detailed in their supplemental materials. By symmetry it is enough to calculate $S^{zz}(\textbf{k}, \omega)$, which we did using the Krylov-space correction vector approach~\cite{kuehner-99, nocera-16A}. Open boundary conditions (OBC) were employed. For finite-temperature DMRG, we applied the ancilla (or purification) method~\cite{nocera-16, feiguin-05, feiguin-10} with a grand-canonical entangler to a two-leg ladder geometry with $L=50$ physical sites along one leg, and 50 ancilla sites along the other leg. We targeted and achieved a truncation error below 10$^{-10}$ by keeping a minimum of 100 and up to 1000 states in the calculations.

The results of the calculated dynamical structure factor $S(\textbf{k}, \omega)$ reported herein use a frequency step $\Delta \omega = 0.02J$ and incorporate a Lorentzian energy broadening, with half width at half maximum $\eta = 0.1J$. We note that the optimal energy broadening in the simulation is limited by the finite size of the system according to $\eta \propto 1/L$~\cite{jeckelmann-02}. With increasing system size and decreasing $\eta$, the dispersion is expected to grow sharper, with a higher intensity maximum. The choice of $\eta = 0.1J \approx 0.38$ meV has previously found to be suitable for this system size, but may suppress intensity compared to the Raman experiment, which has an energy resolution better than 0.12 meV. Finite-temperature spectra corresponding to the experimental temperatures in Fig. 2(c) are shown in Fig. 3(a)-(h), displaying the typical behavior of the Heisenberg antiferromagnetic chain, developing from a diffuse continuum at high temperatures towards increasingly pronounced antiferromagnetic correlations at low temperatures.

\begin{figure*}
\includegraphics[width=16cm]{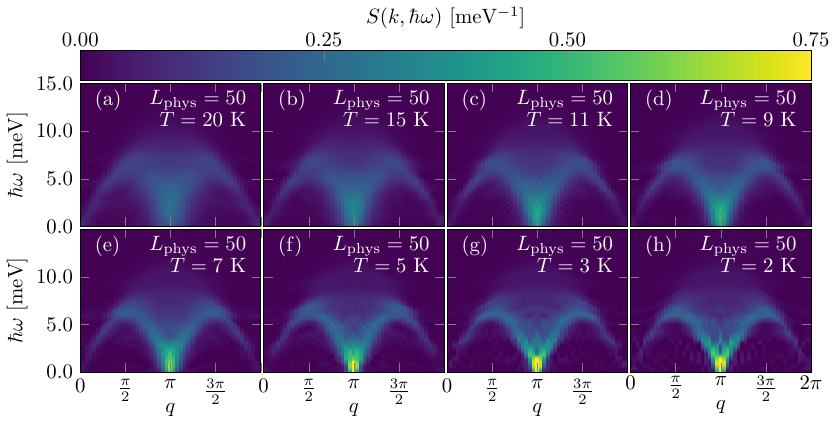}
\caption{\label{fig:DMRG1} \textbf{Finite-temperature dynamical structure factors for $\beta$-VOSO$_4$ as a function of energy $\hbar \omega$ and momentum $q$.} The structure factors $S(\textbf{k},\hbar \omega)$ are calculated for \textbf{a)} $T$ = 20 K, \textbf{b)} $T$ = 15 K, \textbf{c)} $T$ = 11 K, \textbf{d)} $T$ = 9 K, \textbf{e)} $T$ = 7 K, \textbf{f)} $T$ = 5 K, \textbf{g)} $T$ = 3 K, and \textbf{h)} $T$ = 2 K, using DMRG for a chain with $L=50$ sites using open boundary conditions and keeping up to a total of $m_{\mathrm{max}} = 1000$ states. The colour scale represents the weight of the dynamical structure factor at a given momentum and energy.}
\end{figure*}

\subsection{Quantum Fisher Information}

Following Hauke \textit{et al.}~\cite{hauke-16}, the QFI density, $f_\mathrm{Q}$, is related to the dynamical susceptibility, $\chi''$, through

\begin{equation}
f_\mathrm{Q}(\textbf{q},T) = \frac{4}{\pi} \int_0^{\infty}{d(\hbar \omega) \texttt{tanh} \left( \frac{\hbar \omega}{2k_\mathrm{B}T} \right) \chi'' (\textbf{q},\omega,T)},
\end{equation}
where the system is assumed to be at thermal equilibrium at temperature $T$. The dynamical susceptibility is related to a dynamical correlation function through a fluctuation-dissipation theorem relevant to the spectroscopic technique used. In the case of neutron scattering data that has not been frequency-symmetrized, the relevant expression is~\cite{lovesey-86, scheie-25}

\begin{equation}
\chi''(\textbf{q},\omega,\beta) = \pi \left( 1-\texttt{e}^{-\hbar \omega \beta} \right) S(\textbf{q},\omega,\beta),
\end{equation}
where $\beta = 1/(k_\mathrm{B} T)$ is the inverse temperature. To obtain the QFI in absolute units, we normalize it such that the structure factor satisfies the sum rule

\begin{equation}
\sum_{\alpha \in \{x,y,z\}}{\int_{-\infty}^{\infty} \int_{0}^{2\pi}{d\omega d\textbf{k} S^{\alpha \alpha}} = S(S+1)},
\end{equation}
where $S$ is the spin length (here $S = 1/2$ and, by isotropy, $S^{\alpha \alpha} (\textbf{k}, \omega) = S^{zz}(\textbf{k}, \omega))$. Next, in the context of ``regular'', non-polarized INS it is convenient to normalize the QFI according to~\cite{scheie-21}

\begin{equation}
\mathrm{nQFI}(\textbf{q},T) \equiv \frac{f_\mathrm{Q}(\textbf{q},T)}{12S^2},
\end{equation}
where nQFI stands for normalized QFI. When

\begin{equation}
\mathrm{nQFI} > m,
\end{equation}
with $m$ an integer and divisor of the system size, the QFI witnesses at least $(m + 1)$-partite entanglement. Note that this approach can never certify the absence of entanglement in the system, yet it can only confirm its presence.

\begin{figure}
\includegraphics[width=8cm]{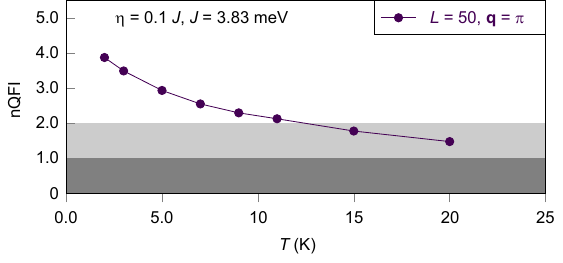}
\caption{\label{fig:DMRG3} \textbf{Normalized quantum Fisher information (nQFI).} The plotted datapoints are obtained from the calculated spectra shown in Fig. 3 for a system of $L = 50$ sites at momentum $\textbf{q}=\pi$ and an exchange interaction $J = 3.83$ meV. Multipartite entanglement is witnessed for $2<\mathrm{nQFI}$. In the light shaded area with $1<\mathrm{nQFI}<2$, at least bipartite entanglement is witnessed. For $\mathrm{nQFI}<1$ (dark shaded area), QFI does not witness (but also does not rule out) entanglement.}
\end{figure}

Using the dynamical spin structure factors plotted in Fig. 3, we obtain the nQFI for $\textbf{q}=\pi$, corresponding to the antiferromagnetic wave vector. The resulting theoretical values are plotted in Fig. 4.

\section{Discussion}

Having established our experimental and theoretical results, we now attempt to correlate the measured Raman spectral weight of the spinon continuum with the calculated QFI. We first obtain the Bose-corrected Raman scattering response $\chi''_\mathrm{R} (\omega,T)$, which is related to the as-measured intensity $I_\mathrm{R} (\omega,T)$ through the fluctuation-dissipation theorem as
\begin{equation}
I_\mathrm{R} (\omega,T) = [1+n(\omega,T)] \cdot \chi''_\mathrm{R} (\omega,T),
\end{equation}
where $n(\omega,T)$ is the Bose factor. Subsequently, following Eq. (2), we apply a \texttt{tanh} filter function to the Bose-corrected data, and integrate the obtained spectral weight from 0 to $\infty$.

In Fig. 5(a), we plot the experimentally obtained QFI integrand for four selected temperatures $T=2$ K, 7 K, 11 K, and 20 K (black lines). To highlight the contribution of spinon excitations we add solid red lines based on a simple damped-oscillator function. While serving as guidelines only, these functions nevertheless approximate the spectral weight of the experimental QFI integrand with high accuracy.

The spin chains in $\beta$-VOSO$_4$ form a zigzag structure with weak interchain coupling, anisotropic magnetic interactions, as well as longer-range exchange interactions~\cite{quintero-castro-22}. In contrast to a uniform nearest-neighbor Heisenberg spin chain, these additional contributions result in a partial confinement of spinon excitations. Within this context, the four-spin correlations can be approximately factorized into products of two-spin correlations, $\langle S_i S_j S_k S_l \rangle \sim \langle S_i S_j \rangle \langle S_k S_l \rangle$. This approximation provides a reasonable connection between the observed magnetic Raman continuum and the dynamical spin structure factor across the Brillouin zone. A rigorous theoretical treatment of magnetic excitations in a buckled spin chain has been recently carried out for the related Ba$_4$Ir$_3$O$_{10}$~\cite{sokolik-22}. Based on these considerations, we note that a direct match between the experimental data and linecuts from our DMRG results taken at specific momenta (e.g., at $\textbf{q}=0$ or $\textbf{q}=\pi$) would be far from obvious. Instead, a full theoretical description requires a nuanced weighting of contributions from momenta at which the local density of states is large. Nevertheless, we can use our simplistic approach to look for similar scaling behavior between the thermal evolution of the magnetic spectral weight probed by Raman spectroscopy and DMRG-based nQFI values. For this final step, we normalize the Raman spectral weight to coincide with the calculated QFI, where the minimal number of entangled particles $n_\mathrm{e}$ is given as $n_\mathrm{e} = m+1$:
\begin{equation}
\mathrm{nQFI}_\mathrm{R} (T)+1 = \frac{f_\mathrm{Q,R}(T)}{12S^2}+1 > m+1 = n_\mathrm{e}.
\end{equation}

\begin{figure*}
\includegraphics[width=12cm]{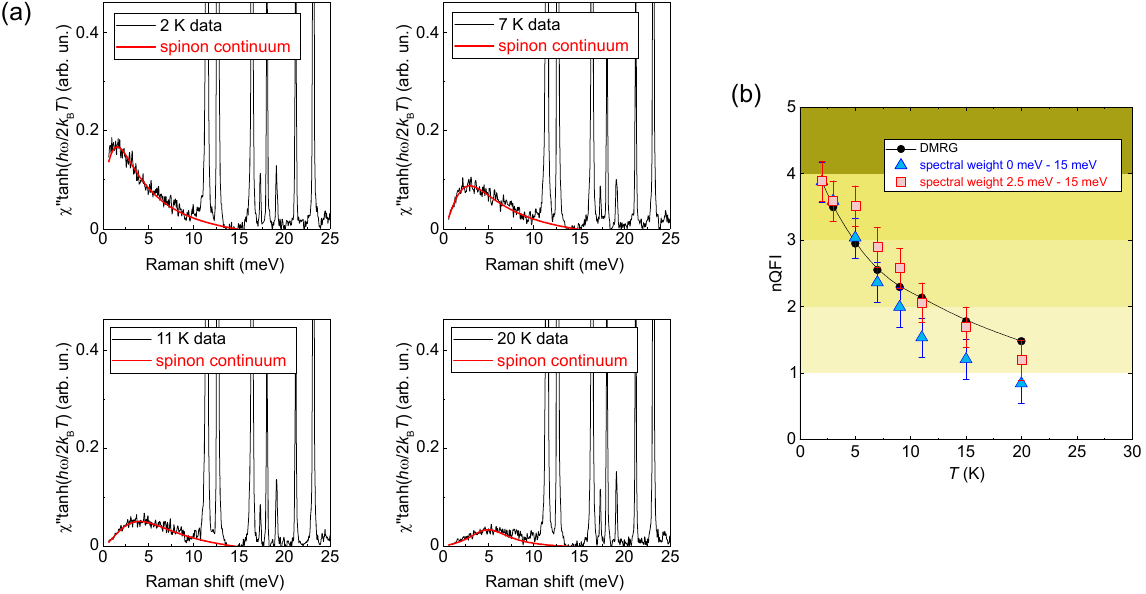}
\caption{\label{fig:nQFI} \textbf{Comparison of Raman-measured $\chi''$ to DMRG results.} \textbf{a)} Normalized quantum Fisher information integrand related to Bose-corrected Raman data $\chi''$ measured at four selected temperatures $T=2$ K, 7 K, 11 K, and 20 K (black lines). Red lines highlight the spectral weight of the spinon continuum. \textbf{b)} Normalized quantum Fisher information (nQFI) as entanglement witness. Blue triangles: Raman-measured spectral weight integrated from 0 meV to 15 meV; red squares: Raman-measured spectral weight integrated from 2.5 meV to 15 meV; black circles: DMRG results shown in Fig. 4. The successively darkening shading along the vertical direction marks the increasing multipartite entanglement. All experimental values have been normalized to the DMRG value at 2 K. The error bars denote the standard deviation obtained from the fitting procedures.}
\end{figure*}

In Fig. 5(b), we trace the thermal evolution of particle entanglement ($n_\mathrm{e}$). As a reference, we use the DMRG-calculated values from Fig. 4 (black circles in Fig. 5). At first, we compare the experimental values integrated over the full spectral range of spinon excitations [i.e., from 0 meV to 15 meV; blue triangles in Fig. 5(b)] to DMRG results. At low temperatures this results in an excellent agreement, while towards higher temperatures a divergence between experimental and theoretical values occurs. This may be rationalized by the fact that the experiment is plagued by increasing thermal fluctuations at finite sample temperature. These fluctuations efficiently suppress higher-order scattering processes from the second half of the Brillouin zone, thus limiting the observable range of magnetic excitations to the first half, which tends to suppress spectral weight at high energies~\cite{loosdrecht-96}. We test this hypothesis by intentionally limiting our integration range to 2.5 meV -- 15 meV (red squares), thereby de-emphasizing low-energy contributions. Indeed, this simple adjustment results in a closer comparison between theory and experiment. 

While this approach yields very satisfactory results, we emphasize an important caveat: Our analysis explicitly relies on the assumptions that the four-spin correlations are approximated as products of two-spin correlations and that the entanglement vanishes around the temperature scale set by $J$ (corresponding to the uncorrelated paramagnetic state). Within this framework, we benchmark DMRG-based QFI calculations to deduce entanglement values from Raman scattering intensities measured in arbitrary units. Our study may therefore serve as a conceptual testing ground for a well-established spin-chain model. In this light, our obtained value of nQFI$_\mathrm{R}$ does not witness at least 4-partite entanglement. Nonetheless, since QFI is conceptually related to spin correlation lengths, our analysis indicates that the spin correlation lengths increase four times when cooling from the paramagnetic state (at $T \sim J$) to the liquid-like state ($T \sim 0.1 J$).

Ultimately, finding an optimized filtering function for the DMRG results that includes the influence of thermal fluctuations at finite temperatures on the high-energy side of the spectral weight may establish a more intimate link between theory and experiment. As such, Raman spectroscopy has the potential to become an important experimental method to witness quantum entanglement, complementary to other recently discussed approaches, such as scanning tunneling microscopy, angular resolved photoemission spectroscopy, and (resonant) inelastic X-ray spectroscopy~\cite{malla-24, liu-25, ren-24, balut-25}. Importantly, Raman spectroscopy allows for an in-situ monitoring of entanglement in quantum magnets under extreme conditions such as strain, pressure, and high magnetic fields, which are oftentimes inaccessible with other experimental techniques.

In summary, we investigated the magnetic excitations and their correlations in the quasi-one-dimensional magnet $\beta$-VOSO$_4$. Fractionalized spinons associated with four-spin correlations were identified, and the QFI extracted from the spinon continuum spectra was compared with finite-temperature DMRG simulations. The resulting scaling alludes towards at least four-times growth of spin correlation lengths at low temperatures, comparable to that observed in other spin chain systems, highlighting the potential effectiveness of a combined DMRG and Raman spectroscopy approach in quantifying spin correlation lengths through QFI.

\section{\label{exp}Methods}

\subsection{Sample Synthesis}
Single crystals of $\beta$-VOSO$_4$ were prepared following the established method~\cite{quintero-castro-22, sieverts-28}: initially, 3 g of V$_2$O$_5$ and 100 ml of H$_2$SO$_4$ were mixed. This solution was heated up to 290$^{\circ}$C at a heating rate of 135$^{\circ}$C/h in a conical flask while connected to a fractionating column where all vapors were condensed. After long periods of time, the solution was cooled down at a rate of 30$^{\circ}$C/h. All products were washed in an ice-water bath and dried at 110$^{\circ}$C for 12 h. $\beta$–VOSO$_4$ samples were the remaining solid parts of the reaction, which yielded green, semi-transparent specimens with typical dimensions of about 5 mm $\times$ 1 mm $\times$ 100 $\mu$m.

\subsection{Raman Spectroscopy}
Temperature- and polarization-resolved Raman scattering experiments were carried out using diode-pumped continuous-wave lasers emitting at 561 nm (Oxxius-LCX) and at 660 nm (Cobolt 05-01 series). The sample was directly mounted onto the cold-finger of an open-cycle He-flow cryostat (Oxford) via silver glue (Ted Pella, Inc.). The laser was focused onto the sample with a spot of about 2 $\mu$m in diameter and a laser power of less than 80 $\mu$W to reduce local laser heating effects. The polarization was selected using a super-achromatic $\lambda/2$-waveplate (Thorlabs) and a linear polarizer as the analyzer. The laser lines were discriminated with volume Bragg grating sets (OptiGrate) which allow a low-energy spectral cut-off at 6 cm$^{-1}$ or less. The inelastically scattered light was dispersed through a single-stage spectrometer (Princeton Instruments HRS-750) and recorded by a nitrogen-cooled charge-coupled device (PyLoN eXcelon).

\begin{acknowledgments}
The work of P.L. and E.D. was supported by the U.S. Department of Energy (DOE), Office of Science, Basic Energy Sciences (BES), Materials Sciences and Engineering Division. G.A. contributed to the DMRG work, and was supported by the U.S. DOE, Office of Science, National Quantum Information Science Research Centers, Quantum Science Center. D.W. was supported by the faculty research fund of Sejong University in 2025, and by the Institute of Applied Physics of Seoul National University. K.Y.C. acknowledges support from the Nano-Material Technology Development Program through the National Research Foundation of Korea (NRF) funded by Ministry of Science and ICT (RS-2023-00281839).
\end{acknowledgments}

\end{document}